\documentclass[11pt]{article}

\usepackage[top=50mm, bottom=50mm, left=50mm, right=50mm]{geometry}

\usepackage{amsmath}
\usepackage{amssymb}
\usepackage{amsfonts}
\usepackage{mathtools}
\usepackage{mathrsfs}
\usepackage[scr = dutchcal]{mathalpha}
\usepackage{graphicx}
\usepackage{graphics}
\usepackage{float}
\usepackage{color}
\usepackage{fullpage}
\usepackage[normalem]{ulem} 
\usepackage{makeidx}
\usepackage{xspace}
\usepackage{authblk}
\usepackage{datetime}
\usepackage{microtype}
\usepackage{xpatch}
\usepackage{xstring}
\usepackage{fancyhdr}
\usepackage{mathrsfs}
\usepackage{esint}
\usepackage{upgreek}
\usepackage{siunitx}
\usepackage{chemformula}
\usepackage{tabularx, multirow}
\usepackage{threeparttable}
\usepackage[]{booktabs} 
\usepackage{float}
\usepackage[section]{placeins}
\usepackage{graphicx,xcolor}
\usepackage[colorlinks=false]{hyperref}
\usepackage[noabbrev,sort&compress]{cleveref}
\usepackage[font = footnotesize, labelfont=bf]{caption}
\usepackage{setspace}
\usepackage{cite}

\fancypagestyle{fpg}{%
	\lfoot{\footnotesize Preprint submitted to Materials }%
	\cfoot{\normalsize\thepage}
	\rfoot{\footnotesize \textbf{2021}\href{}{}}%
}

\makeindex

\author[1,$\dagger$]{Amin Ebrahimi}
\author[1]{Aravind Babu}
\author[2]{Chris R. Kleijn}
\author[1]{Marcel J. M. Hermans}
\author[1]{Ian~M.~Richardson}

\affil[1]{\textit{Department of Materials Science and Engineering, Faculty of Mechanical, Maritime and Materials~Engineering, Delft~University~of~Technology, Mekelweg~2, 2628CD~Delft, The~Netherlands}}
\affil[2]{\textit{Department of Chemical Engineering, Faculty of Applied Sciences, Delft~University~of~Technology, van~der~Maasweg~9, 2629HZ~Delft, The~Netherlands}}
\affil[$\dagger$]{Corresponding author: A.Ebrahimi@tudelft.nl (A. Ebrahimi)}

\title{\Large\textbf{The~Effect of Groove Shape on Molten Metal Flow Behaviour in Gas~Metal~Arc~Welding}}
\date{}

\begin{document}

\maketitle
\thispagestyle{fpg}

\begin{abstract}
		One of the~challenges for development, qualification and optimisation of arc welding processes lies in characterising the~complex melt-pool behaviour which exhibits highly non-linear responses to variations of process parameters. The~present work presents a~simulation-based approach to describe the~melt-pool behaviour in root-pass gas metal arc welding (GMAW). Three-dimensional numerical simulations have been performed using an~enhanced physics-based computational model to unravel the~effect of groove shape on complex unsteady heat and fluid flow in GMAW. The~influence of surface deformations on power-density distribution and the~forces applied to the~molten material were taken into account. Utilising this model, the~complex heat and fluid flow in melt pools was visualised and described for different groove shapes. Additionally, experiments were performed to validate the~numerical predictions and the~robustness of the~present computational model is demonstrated. The~model can be used to explore physical effects of governing fluid flow and melt-pool stability during gas metal arc root welding.
\end{abstract}

\bigskip
\newpage

\onehalfspacing
\section{Introduction}
\label{sec:intro}
Gas metal arc welding (GMAW) is a~fusion-based joining technique that is widely employed in industry to join metallic parts and to produce high-integrity structures. The~quality of the~joints made using arc welding or the~structures made using wire-arc additive manufacturing depend on chosen process parameters, material properties and boundary conditions~\cite{Wei_2021,Norrish_2021,DebRoy_1995}. Changes in operating variables can alter the~magnitude and distribution of the~heat input and forces applied to the~molten metal in melt pools (such as Marangoni, Lorentz, thermal buoyancy forces and arc plasma shear stresses and pressures), affecting fluid flow in the~pool and in turn the~properties, structure and quality of products~\cite{Norrish_2021}. Correct control of melt-pool behaviour during arc welding is crucial to produce joints with desired properties~\cite{Aucott_2018}.

One of the~challenges for development, qualification and optimisation of arc welding processes lies in characterising the~complex melt-pool behaviour which exhibits highly non-linear responses to variations of process parameters~\cite{David_1992}. Trial-and-error experiments are often employed to realise appropriate processing parameters to achieve the~desired properties. Such an~experimental approach is costly and time inefficient and a~successful processing for a~specific configuration (\textit{e.g.}~material system, welding machine and joint shape) might not apply to a~different configuration. Moreover, experimental identification of the~effects of various parameters on the~melt-pool behaviour is generally complicated due to the~high-temperature, rapid solid-liquid phase transformation, opacity and fast dynamics of the~molten metal flow~\cite{Aucott_2018}. Simulation-based approaches offer understanding of the~melt-pool behaviour during welding and additive manufacturing and can serve as an~alternative to experiments to explore the~design space for process optimisation~\cite{Wei_2021,Cook_2020}.

To date, focus has predominantly been placed on developing numerical simulations to describe melt-pool behaviour in arc welding of flat plates without a~groove (\textit{i.e.} bead-on-plate welding, see for instance, \cite{Zong_2021,Hu_2021,Zargari_2020,DWu_2020,Wu_2017,Hu_2008,Cho_2007,Ushio_1997,Kim_1994}); however little attention has been paid to understanding the~effect of joint shape on complex heat and molten metal flow. \mbox{Zhang~\textit{et~al.}~\cite{Zhang_2004_a,Zhang_2004_b}} developed a~three-dimensional model in a~body-fitted coordinate system to describe the~effects of various driving forces on heat and fluid flow in the~melt pool during GMAW fillet welding. \mbox{Hu~and~Tsai\cite{Hu_2008_b}} developed a~comprehensive model to simulate unsteady molten metal flow and heat transfer in melt-pools during GMA welding of a~thick plate with V-groove. These studies only focus on partially penetrated pools and do not report the~effect of different joint shapes on molten metal flow behaviour. Chen~\textit{et~al.}~\cite{Chen_2011} studied the~effect of the~opening angle of a~V-groove on melt-pool behaviour during relatively high-current GMAW (welding current $I = \SI{340}{\ampere}$) using a~computational model developed on the~basis of a~body-fitted coordinate system. They reported that changes in the~opening angle have an~insignificant effect on the~flow pattern in the~pool but can affect the~velocity and temperature distribution and thus the~pool shape. Using the~Abel inversion method, Cho~and~Na~\cite{Cho_2005} reconstructed the~emissivity distribution of an~arc plasma and argued that the~application of V-grooves in arc welding can affect the~arc plasma characteristics, changing the~distribution of the~power-density, arc pressure and electromagnetic forces~\cite{Cho_2013}. On the~basis of their previous studies~\cite{Cho_2005,Cho_2013}, \mbox{Cho~\textit{et~al.}~\cite{Cho_2013_b}} employed an~elliptically symmetric distribution functions for power-density and arc pressure (instead of an~axisymmetric distribution) to simulate heat and fluid flow in GMAW of a~plate with V-groove at different welding positions. Changes in the~groove shape due to filler metal deposition and its effect on the~distribution of power-density and arc-induced forces were not accounted for in previous models that are available in the~literature. Further investigations are required to realise the~influence of the~joint shape on molten metal flow behaviour in GMAW, particularly for fully-penetrated melt pools. 

Focusing on understanding the~melt-pool behaviour during root-pass gas metal arc welding, with particular interest in the~effects of groove shape, a~systematic numerical study was carried out in the~present work. Three-dimensional calculations have been performed using a~physics-based computational model to simulate the~dynamics of heat and molten metal flow in GMAW. Additionally, experiments were performed to validate the~numerical predictions. The~present work explains the~dynamics of internal molten metal flow in gas metal arc welding and provides an~enhanced computational model for design space explorations.
\FloatBarrier

\section{Problem description}
\label{sec:problem_des}
\begin{figure}[!htb] %[H] %[htbp]
	\centering
	\includegraphics[width=\linewidth]{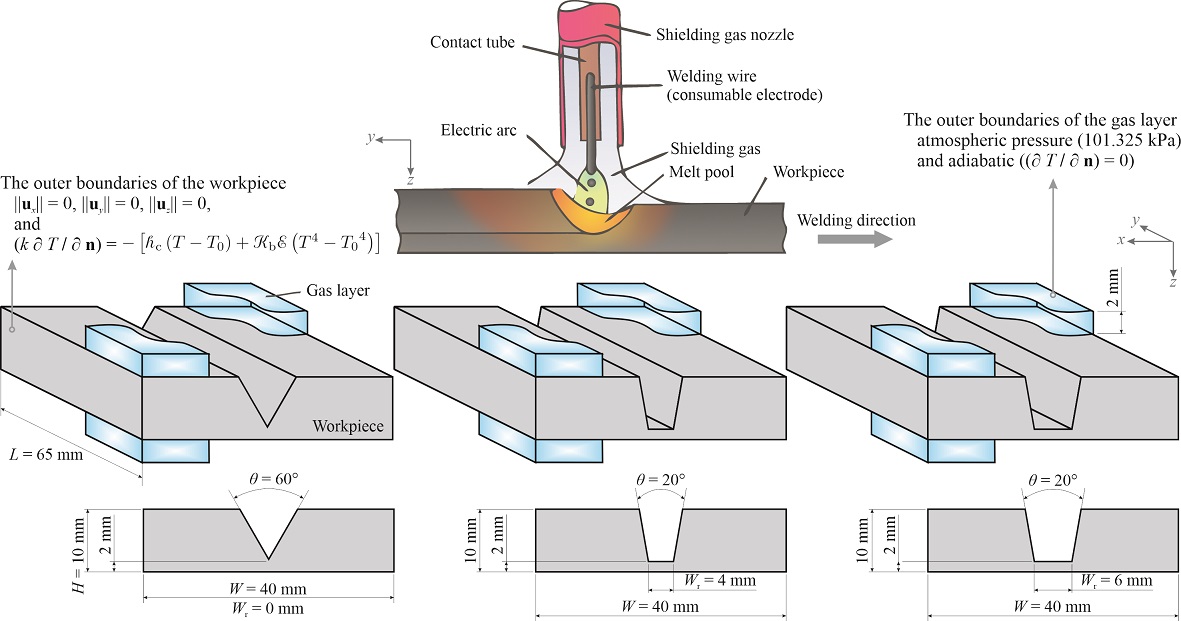}
	\caption{Schematic of gas metal arc welding and three different joint shapes studied in the~present work. For the~sake of clarity, only parts of the~gas layer are shown. Here $W_\mathrm{r}$ refers to the~width of flat region at the~base of the~groove, referred to as the~root-leg.}
	\label{fig:schematic}
\end{figure}

In gas metal arc welding, an~electric arc between a~consumable electrode (filler metal) and a~workpiece provides the~thermal energy required for melting the~material. Melting of the~filler metal results in the~periodic formation of molten metal droplets that successively impinge on the~workpiece surface. Thermal energy input from the~arc plasma as well as the~thermal energy transported by the~droplets leads to the~formation of a~melt pool that creates a~joint after solidification (see~\cref{fig:schematic}). In the~present work, the~effect of the~groove shape on molten metal flow behaviour is studied for three different groove shapes, as shown schematically in~\cref{fig:schematic}. A~torch, which is perpendicular to the~workpiece top-surface is adopted here and the~contact-tip to workpiece distance (CTWD) is set to $\SI{18}{\milli\meter}$. Different values of welding current ranging between $\SI{220}{\ampere}$ and $\SI{280}{\ampere}$ have been studied. Details of the~welding parameters in the~present work are listed in \cref{tab:welding_parameters}. The~process parameters employed in the~present work have been chosen based on preliminary trial experiments and are also comparable to those reported in previous independent studies on gas metal arc welding of steel plates with grooves (see for instance, \cite{Cho_2013_b,Cho_2013,Hu_2008_b,Zhang_2004_b}). The~plates are made of a~stainless steel alloy~(AISI~316L) and are initially at an~ambient temperature of $\SI{300}{\kelvin}$. The~welding torch is initially located in the~middle of the~workpiece along the~$x$-axis and \SI{10}{\milli\meter} away from the~leading-edge of the~workpiece (\textit{i.e.} $y = \SI{10}{\milli\meter}$).

\bgroup
\def\arraystretch{1.0}	
\begin{table}[!htb] %[H] %[htbp]
	\centering
	\caption{Welding parameters studied in the~present work.}
	\small
	\begin{tabular}{lll}
		\toprule
		Parameter                           & Value                           & Unit                          \\ \midrule
		Welding current $I$                 & \SI{220}-\SI{280}               & \si{\ampere}                  \\
		Arc voltage $U$                     & \SI{21.4}-\SI{23.0}             & \si{\volt}                    \\
		Wire feed rate $u_\mathrm{w}$       & \SI{7.0}-\SI{8.7}               & \si{\meter\per\minute}        \\
		Wire diameter $d_\mathrm{w}$        & \SI{1.2} (\SI{0.045})           & \si{\milli\meter} (\si{inch}) \\
		Wire material                       & AISI~316L                       & --                            \\
		Travel speed $V$                    & \SI{7.5}                        & \si{\milli\meter\per\second}  \\
		Shielding gas                       & 97.5\%~\ch{Ar} + 2.5\%~\ch{CO2} & --                            \\
		Shielding gas flow rate             & \SI{20}                         & \si{\litre\per\minute}        \\
		Inner diameter of the~shielding cup & \SI{20}                         & \si{\milli\meter}             \\
		CTWD                                & \SI{18}                         & \si{\milli\meter}             \\
		Distance between the~contact tip    & \SI{2}                          & \si{\milli\meter}             \\
		and the~shielding cup edge          &                                 &                               \\
		Torch angle                         & \SI{90}                         & $^{\circ}$                    \\ \bottomrule
	\end{tabular} 
	\label{tab:welding_parameters}
\end{table}
\egroup	

\begin{figure}[!htb] %[H] %[htbp]
	\centering
	\includegraphics[width=\linewidth]{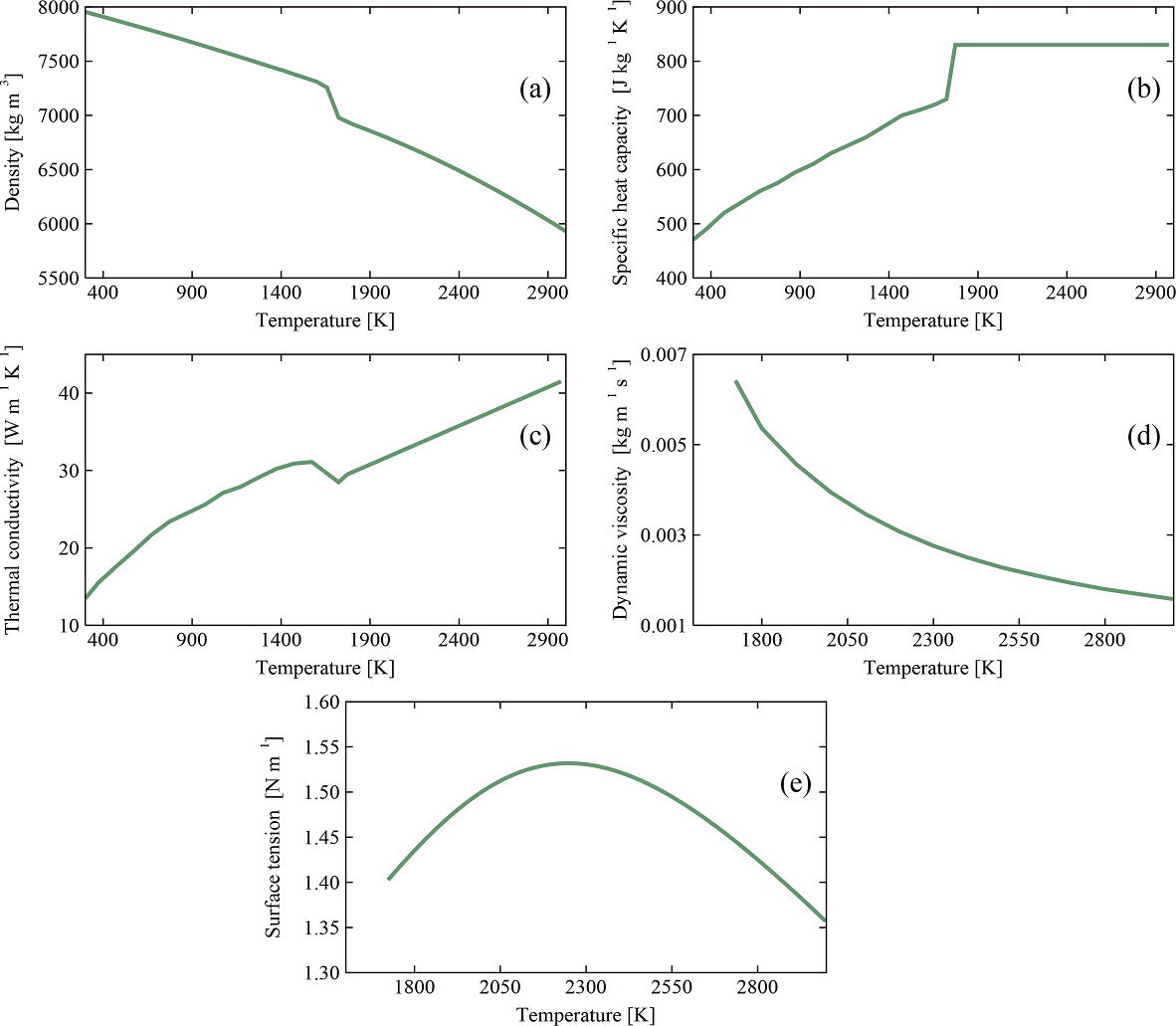}
	\caption{Temperature-dependent thermophysical properties of AISI~316L employed in the~simulations. (a)~density~\cite{Kim_1975}, (b)~specific heat capacity at constant pressure~\cite{Mills_2002_316}, (c)~thermal conductivity~\cite{Mills_2002_316}, (d)~dynamic viscosity~\cite{Kim_1975} and (e)~surface~tension~\cite{Sahoo_1988}.}
	\label{fig:materials_properties}
\end{figure}

The~computational domain is defined in a~stationary Cartesian coordinate system and is in the~form of a~rectangular cube that encompasses the~metallic workpiece and two layers of gas below and above the~workpiece. The~incorporation of the~gas layers allows tracking of surface deformations of the~pool. To reduce the~complexity of simulations and computation time, the~melt-pool is decoupled from the~arc plasma in the~simulations. Accordingly, the~heat input from the~arc and the~arc induced forces are defined as source terms for thermal energy and momentum. These source terms are adjusted dynamically during the~calculations, as explained in~\cref{sec:methods}, to account for the~changes in the~arc power and power-density distribution as well as the~magnitude and distribution of the~forces exerted by the~arc plasma that occur due to melt-pool surface deformations and filler metal deposition. The~conditions applied to the~outer boundaries of the~computational domain are shown in~\cref{fig:schematic}. The~outer boundaries of the~plates are treated as no-slip walls, as the~melt-pool does not reach them, and heat losses due to radiation and convection are accounted for. A~fixed atmospheric pressure ($\SI{101325}{\pascal}$) is applied to the~outer boundaries of the~gas layers. The~thermophysical properties of AISI~316L and the~gas employed in the~simulations are presented in \cref{tab:material_properties} and \cref{fig:materials_properties}. The~values for the~surface tension are estimated using an~empirical correlation proposed by Sahoo~\textit{et~al.}~\cite{Sahoo_1988}, which takes the~influence of surfactants (\textit{i.e.} sulphur) into account. Employing a~temperature-dependent density model, thermal buoyancy force are accounted for in the~simulations. In the~present work, the~properties of the~shielding gas are assumed to be temperature-independent, which is a~common assumption in numerical simulations of arc welding and additive manufacturing where the~melt-pool is decoupled from the~arc plasma~\cite{Cho_2013_b,Cho_2007,Hu_2008,Wu_2017,Zargari_2020,Hu_2021,Zong_2021}. This assumption is justifiable as the~transport properties of the~shielding gas (\textit{i.e.} viscosity, density and thermal conductivity) are small compared to those of the~molten metal, and thus changes in the~shielding gas properties with temperature negligibly affect the~numerical predictions of fluid flow in the~melt pool~\cite{Saldi_2012_thesis}.

\bgroup
\def\arraystretch{1.0}	
\begin{table}[!htb] %[H] %[htbp]
	\centering
	\caption{Thermophysical properties of the~stainless steel (AISI~316L) and the~gas employed in the~numerical simulations. Values for AISI~316 are taken from~\cite{Mills_2002_316}.}
	\small
	\begin{tabular}{llll}
		\toprule
		Property                              & Stainless steel (AISI~316)          & Gas           & Unit                                \\ \midrule
		Density $\rho$                        & see \cref{fig:materials_properties} & \SI{1.623}    & \si{\kilogram\per\meter\cubed}      \\
		Specific heat capacity $c_\mathrm{p}$ & see \cref{fig:materials_properties} & \SI{520.64}   & \si{\joule\per\kilogram\per\kelvin} \\
		Thermal conductivity $k$              & see \cref{fig:materials_properties} & \SI{1.58e-2}  & \si{\watt\per\meter\per\kelvin}     \\
		Viscosity $\mu$                       & see \cref{fig:materials_properties} & \SI{2.12e-05} & \si{\kilogram\per\meter\per\second} \\
		Latent heat of fusion $\mathcal{L}$   & \SI{2.7e5}                          & --            & \si{\joule\per\kilogram}            \\
		Liquidus temperature $T_\mathrm{l}$   & \SI{1723}                           & --            & \si{\kelvin}                        \\
		Solidus temperature $T_\mathrm{s}$    & \SI{1658}                           & --            & \si{\kelvin}                        \\ \bottomrule
	\end{tabular} 
	\label{tab:material_properties}
\end{table}
\egroup	

\FloatBarrier

\section{Methods} 
\label{sec:methods}

\subsection{Mathematical model}
\label{sec:math_model}	
A~three-dimensional multiphase model has been developed to predict molten metal flow, heat transfer and associated surface movements in gas metal arc welding. In~the~present~model, the~fluids (\textit{i.e.} the~molten metal and the~gas) are considered to be Newtonian and their densities are assumed to pressure-independent. Assuming that the~fluid flows under consideration are in the~continuum regime, the~dynamics of heat and fluid flow in melt pools and their surroundings are governed by the~equations of motion given by the~conservation equations for mass, momentum and energy. Accordingly, the~unsteady governing equations are cast as follow:

\begin{equation}
	\frac{D \rho}{D t} = S_\mathrm{m},
	\label{eq:mass}
\end{equation}

\begin{equation}
	\rho \frac{D \mathbf{u}}{D t} = \mu \nabla^2 \mathbf{u} -\nabla p + \mathbf{F}_\mathrm{d} + \mathbf{F}_\mathrm{s} + \mathbf{F}_\mathrm{b} + S_\mathrm{m} \left(\mathbf{\mathbf{u}_\mathrm{s}} - \mathbf{u}\right),
	\label{eq:momentum}
\end{equation}

\begin{equation}
	\rho \frac{D h}{D t} = \frac{k}{c_\mathrm{p}} \nabla^2 h - \rho \frac{D \left( \psi \mathcal{L}_\mathrm{f} \right)}{D t} + S_\mathrm{q} + S_\mathrm{l} + S_\mathrm{m} \left({\mathcal{L}_\mathrm{f}} + \int_{T_\mathrm{i}}^{T_\mathrm{d}} {c_\mathrm{p}} \mathop{}\!\mathrm{d}T\right),
	\label{eq:energy}
\end{equation}	

\noindent
where,~$\rho$ is the~density, $\mathbf{u}$~the~relative fluid-velocity vector, $\mathbf{u}_\mathrm{s}$~the~fluid-velocity vector for the~filler metal droplet, $t$~the~time, $\mu$~the~dynamic viscosity, $p$~the~pressure, $h$~the~sensible heat, $k$~the~thermal conductivity, $c_\mathrm{p}$~the~specific heat capacity at constant pressure, $\left( \psi \mathcal{L}_\mathrm{f} \right)$~the~latent~heat, and $S_\mathrm{m}$~the~source term defined to model filler metal addition~\cite{Cho_2006}. The~subscripts `d' and `i' indicate the~droplet and initial condition respectively. The~total enthalpy of the~material~$\mathscr{H}$ is the~sum of the~latent heat~$\left( \psi \mathcal{L}_\mathrm{f} \right)$ and the~sensible heat~$h$ and is defined as follows~\cite{Voller_1991}:

\begin{equation}
	\mathscr{H} = \left(h_\mathrm{r} + \int_{T_\mathrm{r}}^{T} c_\mathrm{p} \mathop{}\!\mathrm{d}T\right) + \psi \mathcal{L}_\mathrm{f},
	\label{eq:enthalpy}
\end{equation}

\noindent
where, $T$~is the~temperature, $\psi$~the~local liquid volume-fraction, and $\mathcal{L}_\mathrm{f}$~the~latent heat of fusion. The~subscript `r' indicates the~reference condition. Assuming the~liquid volume-fraction~$\psi$ to be a~linear function of temperature~\cite{Voller_1991}, its value can be calculated as follows:

\begin{equation}
	\psi = \frac{T - T_\mathrm{s}}{T_\mathrm{l} - T_\mathrm{s}}; \quad T_\mathrm{s} \le T \le T_\mathrm{l},
	\label{eq:liquid_fraction}
\end{equation}

\noindent
where, $T_\mathrm{l}$ and $T_\mathrm{s}$ are the~liquidus and solidus temperatures, respectively.

To capture the~position of the~gas-metal interface, the~volume-of-fluid (VOF) method~\cite{Hirt_1981} is adopted, where the~scalar function~$\phi$ indicates the~local volume-fraction of a~phase in a~given computational cell. The~value of~$\phi$ varies from 0 in the~gas phase to 1 in the~metal phase, and cells with~$0 < \phi < 1$ represent the~gas-metal interface. The~linear advection equation describes the~advection of the~scalar function~$\phi$ as follows: 

\begin{equation}
	\frac{D \phi}{D t} = \frac{S_\mathrm{m}}{\rho}.
	\label{eq:vof}
\end{equation}

\noindent
Accordingly, the~effective thermophysical properties of the~material in each computational cell are determined as follows:

\begin{equation}
	\xi = \phi \, \xi_\mathrm{m} + \left(1-\phi\right) \xi_\mathrm{g},
	\label{eq:mixture_model}
\end{equation}

\noindent
where,~$\xi$ corresponds to thermal conductivity~$k$, specific heat capacity~$c_\mathrm{p}$, viscosity~$\mu$ or density~$\rho$, and subscripts `g' and `m' indicate gas or metal respectively.

Solid--liquid phase transformation occurs in the~temperature range between~$T_\mathrm{s}$ and~$T_\mathrm{l}$ in the~so-called `mushy zone'. To model the~damping of liquid velocities in the~mushy zone, and suppression of liquid velocities in solid regions, the~sink term~$\mathbf{F}_\mathrm{d}$ based on the~enthalpy-porosity technique~\cite{Voller_1987}, is incorporated into the~momentum equation and is defined as

\begin{equation}
	\mathbf{F}_\mathrm{d} = -C\ \frac{(1 - \psi)^2}{\psi^3 + \epsilon} \ \mathbf{u},
	\label{eq:sink_term}
\end{equation}

\noindent
where,~$C$ is the~mushy-zone constant and~$\epsilon$ is a~constant, equal to~$10^{-3}$, employed to avoid division by zero. Depending on the~melting temperature range as well as the~imposed boundary conditions, the~value of the~mushy-zone constant can affect the~numerical predictions of solidification and melting simulations. The~value of the~mushy-zone constant should be assigned appropriately to avoid numerical artefacts in simulations of solid--liquid phase transformations, which is discussed in detail in~\cite{Ebrahimi_2019}. In~the~present work, the~value of the~mushy-zone constant~$C$ was chosen to equal $10^7\,\SI{}{\kilogram\per\square\meter\per\square\second}$~\cite{Ebrahimi_2019}.

To model forces acting on the~gas-metal interface such as surface tension, thermocapillary and arc plasma forces, the~continuum surface force (CSF) model~\cite{Brackbill_1992} is employed. In~the~CSF model, surface forces are considered as volumetric forces acting on the~material contained in grid cells in the~interface region. The~source term~$\mathbf{F}_\mathrm{s}$ is included in \cref{eq:momentum} as follows:

\begin{equation}
	\mathbf{F}_\mathrm{s} = \mathbf{f}_\mathrm{s} \lVert \nabla\phi \rVert \frac{2\rho}{\rho_\mathrm{m} + \rho_\mathrm{g}},
	\label{eq:CSF_model}
\end{equation}

\noindent
where, subscripts `g' and `m' indicate gas or metal respectively. In~\cref{eq:CSF_model},~$\mathbf{f}_\mathrm{s}$ is the~surface force applied to a~unit area, and the~term~$2\rho/\!\left(\rho_\mathrm{m} + \rho_\mathrm{g}\right)$ is employed to abate the~effect of the~large metal-to-gas density ratio by redistributing the~volumetric surface-forces towards the~metal phase (\textit{i.e.}~the~heavier phase). In addition to surface forces, body forces (\textit{i.e.} electromagnetic forces) are incorporated in the~source term~$\mathbf{F}_\mathrm{b}$ in~\cref{eq:momentum}.

To model the~thermal energy input to the~material, the~source~$S_\mathrm{q}$ is included in~\cref{eq:energy}. Moreover, heat losses from the~workpiece surface due to convection and radiation are accounted for by including the~sink term~$S_\mathrm{l}$ in~\cref{eq:energy}.

In gas metal arc welding, the~surface force acting on the~gas-metal interface~$\mathbf{f}_\mathrm{s}$ includes an~arc plasma term, surface tension and thermocapilary forces, and is defined as follows:

\begin{equation}
	\mathbf{f}_\mathrm{s} = \mathbf{f}_\mathrm{a} + \gamma \kappa \hat{\mathbf{n}} + \frac{\mathrm{d} \gamma}{\mathrm{d} T} \left[\nabla T - \hat{\mathbf{n}}\left(\hat{\mathbf{n}} \cdot \nabla T\right)\right],
	\label{eq:surface_force_gma}
\end{equation}

\noindent
where,~$\mathbf{f}_\mathrm{a}$ is arc plasma force,~$\gamma$ the~surface tension,~$\hat{\mathbf{n}}$ the~surface unit normal vector~\mbox{($\hat{\mathbf{n}} = \nabla\phi / \lVert \nabla\phi \rVert$)} and~$\kappa$ the~surface curvature~($\kappa = \nabla\cdot\hat{\mathbf{n}}$).

The~arc plasma force~$\mathbf{f}_\mathrm{a}$ defined in \cref{eq:surface_force_gma} comprises arc plasma shear stress~$\mathbf{f}_\tau$ and arc pressure~$\mathbf{f}_\mathrm{p}$,

\begin{equation}
	\mathbf{f}_\mathrm{a} = \mathbf{f}_\tau + \mathbf{f}_\mathrm{p}.
	\label{eq:arc_force_gma}
\end{equation}

\noindent
The~arc plasma shear stress~$\mathbf{f}_\tau$, which acts at a~tangent to the~surface, is defined as follows~\cite{Bai_2018}:

\begin{equation}
	\mathbf{f}_\tau = \left[\tau_\mathrm{max} \: g_\tau\left(\mathscr{R}, \sigma_\tau \right)\right] \hat{\mathbf{t}},
	\label{eq:arc_shear_force_gma}
\end{equation}

\noindent
where, the~maximum arc shear stress~$\tau_\mathrm{max}$~\cite{Lee_1996,Lee_1995}, the~arc shear stress distribution function~$g_\tau$~\cite{Unnikrishnakurup_2017} and the~surface unit tangent vector~$\hat{\mathbf{t}}$~\cite{Bai_2018} were defined as follows:

\begin{equation}
	\tau_\mathrm{max} = \SI{7e-2}{} I^{1.5} \exp\left(\frac{\SI{-2.5e4}{} \bar{\ell}}{I^{0.985}} \right),
	\label{eq:max_shear_force_gma}
\end{equation}

\begin{equation}
	g_\tau\left(\mathscr{R}, \sigma_\tau \right) = \sqrt{\frac{\mathscr{R}}{{\sigma_\tau}}} \exp\left(\frac{-\mathscr{R}^2}{{\sigma_\tau}^2}\right) ,
	\label{eq:dist_shear_force_gma}
\end{equation}

\begin{equation}
	\hat{\mathbf{t}} = \frac{\mathbf{r} - \hat{\mathbf{n}} \left(\hat{\mathbf{n}} \cdot \mathbf{r}\right)}{\lVert \mathbf{r} - \hat{\mathbf{n}} \left(\hat{\mathbf{n}} \cdot \mathbf{r}\right) \rVert}.
	\label{eq:tangent_vec_gma}
\end{equation}

\noindent
Here,~$I$ is the~welding current in Amperes,~$\bar{\ell}$ the~mean arc length in meters,~$\mathscr{R}$~the~radius in~$x$-$y$ plane (\textit{i.e.}~$\mathscr{R} = \sqrt{x^2 + y^2}$) in meters, and~$\mathbf{r}$ the~position vector in the~$x$-$y$ plane in meters. The~distribution parameter~$\sigma_\tau$ (in meters) is assumed to be a~function of the~mean arc length~$\bar{\ell}$ and current~$I$ and was approximated on the~basis of the~data reported by~\mbox{Lee~and~Na~\cite{Lee_1996}}:

\begin{equation}
	\sigma_\tau = \SI{1.387e-3}{} + I^{-0.595} \bar{\ell}^{0.733}.
	\label{eq:dist_shear_stress_gma}
\end{equation}

\noindent
The~arc pressure~$\mathbf{f}_\mathrm{p}$ is determined as follows~\cite{Lin_1986}:

\begin{equation}
	\mathbf{f}_\mathrm{p} = \mathscr{F}_\mathrm{p} \left[\frac{\mu_0 I}{4 \pi} \frac{I}{2 \pi {\sigma_\mathrm{p}}^2} \exp\left(\frac{-\mathscr{R}^2}{2 {\sigma_\mathrm{p}}^2}\right) \right] \hat{\mathbf{n}},
	\label{eq:arc_pressure_gma}
\end{equation}

\begin{sloppypar}
	\noindent
	where, $I$ is the~current in Ampere, and $\mu_0$~is the~vacuum permeability equal to \mbox{$4\pi \cdot 10^{-7} \, \SI{}{\henry\per\meter}$}. The~distribution parameter~$\sigma_\mathrm{p}$ (in metres) was determined using the~experimental data reported by~Tsai~and~Eagar~\cite{Tsai_1985} as follows:
\end{sloppypar}

\begin{equation}
	\sigma_\mathrm{p} = \SI{7.03e-2}{} \ell^{0.823} + \SI{2.04e-4}{} I^{0.376},
	\label{eq:dist_pressure_gma}
\end{equation}

\noindent
where,~$\ell$~is the~local arc length in meters, and $I$~the~current in Amperes. Changes in surface morphology can cause the~total arc force applied to the~melt-pool surface~($\iiint \limits_{\forall} \lVert \mathbf{f}_\mathrm{p} \rVert \mathop{}\!\mathrm{d}V$) to differ from the~expected arc force ($\mu_0 I^2 / 4 \pi$) due to changes in~$\lVert \nabla\phi \rVert$~\cite{Ebrahimi_2021,Ebrahimi_2021_b}. This numerical artefact is negated by incorporating~$\mathscr{F}_\mathrm{p}$, defined as follows:

\begin{equation}
	\mathscr{F}_\mathrm{p} = \mathscr{j} \, \frac{\mu_0 I^2}{4 \pi} \frac{1}{\iiint \limits_{\forall} \lVert \mathbf{f}_\mathrm{p} \rVert \mathop{}\!\mathrm{d}V}.
	\label{eq:adj_pressure_gma}
\end{equation}

\begin{sloppypar}
	\noindent
	The~dimensionless factor~$\mathscr{j}$ is employed, as suggested by~Lin~and~Eagar~\cite{Lin_1986} and~\mbox{Liu~\textit{et~al.}~\cite{Liu_2015}}, to match the~theoretically determined arc pressure with experimentally measured values, and is calculated as follows:
\end{sloppypar}

\begin{equation}
	\mathscr{j} = 3 + \SI{8e-3}{} I,
	\label{eq:adj_pressure_coeff_gma}
\end{equation}

\noindent
with~$I$ the~welding current in Amperes.

$\mathbf{F}_\mathrm{b}$ in \cref{eq:momentum} is the~body force, which comprises electromagnetic and gravity forces. The~electromagnetic force was computed using the~model proposed by~Tsao~and~Wu~\cite{Tsao_1988} transformed into a~body-fitted coordinate system. Hence, the~body forces are defined as follows:

\begin{equation}
	{\mathbf{f}_\mathrm{b}}_x = \frac{- \mu_0 I^2}{4\pi^2 {\sigma_\mathrm{e}}^2 \mathscr{R}} \exp\left(\frac{- \mathscr{R}^2}{2{\sigma_\mathrm{e}}^2}\right) \left[1 - \exp\left(\frac{- \mathscr{R}^2}{2{\sigma_\mathrm{e}}^2}\right)\right] \left(1 - \frac{z - z^\prime}{H_\mathrm{m} - z^\prime}\right)^2 \left(\frac{x}{\mathscr{R}}\right),
	\label{eq:body_force_x_gma}
\end{equation}

\begin{equation}
	{\mathbf{f}_\mathrm{b}}_y = \frac{- \mu_0 I^2}{4\pi^2 {\sigma_\mathrm{e}}^2 \mathscr{R}} \exp\left(\frac{- \mathscr{R}^2}{2{\sigma_\mathrm{e}}^2}\right) \left[1 - \exp\left(\frac{- \mathscr{R}^2}{2{\sigma_\mathrm{e}}^2}\right)\right] \left(1 - \frac{z - z^\prime}{H_\mathrm{m} - z^\prime}\right)^2 \left(\frac{y}{\mathscr{R}}\right),
	\label{eq:body_force_y_gma}
\end{equation}

\begin{equation}
	{\mathbf{f}_\mathrm{b}}_z = \frac{- \mu_0 I^2}{4\pi^2 {\mathscr{R}}^2 H_\mathrm{m}} \left[1 - \exp\left(\frac{- \mathscr{R}^2}{2{\sigma_\mathrm{e}}^2}\right)\right]^2 \left(1 - \frac{z - z^\prime}{H_\mathrm{m} - z^\prime}\right) + \rho \mathbf{g}.
	\label{eq:body_force_z_gma}
\end{equation}

\noindent
Here, the~distribution parameter for the~electromagnetic force~$\sigma_\mathrm{e}$ is the~same as~$\sigma_\mathrm{p}$, according to \mbox{Tsai~and~Eagar~\cite{Tsai_1985}}, $z^\prime$~is the~position of the~melt-pool surface in $x$-$y$ plane at a~given time $t$, and $\mathbf{g}$~the~gravitational acceleration vector. It should be noted that the~current-density profile is assumed to be Gaussian in the~model proposed by~Tsao~and~Wu~\cite{Tsao_1988} to compute the~electromagnetic forces. Further studies are required to develop a~generic model to approximate the~evolution of current-density profile during gas metal arc welding~\cite{Rao_2010,Xu_2009,Schnick_2010}.

The~thermal energy provided by the~arc is modelled by adding the~source term~$S_\mathrm{q}$ to the~energy equation (\cref{eq:energy}) and was defined as

\begin{equation}
	S_\mathrm{q} = \mathscr{F}_\mathrm{q} \left[\frac{\eta_\mathrm{a} I U}{2 \pi {\sigma_\mathrm{q}}^2} \exp\left(\frac{-\mathscr{R}^2}{2 {\sigma_\mathrm{q}}^2}\right) \lVert \nabla\phi \rVert \frac{2\, \rho \, c_\mathrm{p}}{(\rho \, c_\mathrm{p})_\mathrm{m} + (\rho \, c_\mathrm{p})_\mathrm{g}} \right],
	\label{eq:arc_heat_gma}
\end{equation}

\noindent
where, the~arc efficiency~$\eta_\mathrm{a}$ is defined as follows:

\begin{equation}
	\eta_\mathrm{a} = \eta_\mathrm{p} - \eta_\mathrm{d}.
	\label{eq:efficiency_gma}
\end{equation}

\noindent
Here, $\eta_\mathrm{p}$ is the~process efficiency and is assumed to vary linearly with welding current from~$77\%$ at~$\SI{200}{\ampere}$ to~$72\%$ at~$\SI{300}{\ampere}$~\cite{Murphy_2018}, and $\eta_\mathrm{d}$~is the~efficiency of thermal energy transfer by molten metal droplets, which is defined as follows:

\begin{equation}
	\eta_\mathrm{d} = \frac{q_\mathrm{d}}{IU},
	\label{eq:drop_efficiency_gma}
\end{equation}

\noindent
with $q_\mathrm{d}$~the~thermal energy content of the~droplets that are assumed to be spherical. $q_\mathrm{d}$~is defined as follows:

\begin{equation}
	q_\mathrm{d} = \rho_\mathrm{d} \frac{4}{3} \pi r_\mathrm{d}^3 \left({\mathcal{L}_\mathrm{f}} + \int_{T_\mathrm{i}}^{T_\mathrm{d}} {c_\mathrm{p}} \mathop{}\!\mathrm{d}T\right) f_\mathrm{d},
	\label{eq:drop_heat_gma}
\end{equation}

\noindent
where, $r_\mathrm{d}$~is the~radius of molten metal droplet. The~droplet temperature $T_\mathrm{d}$ was approximated to~$\SI{2500}{\kelvin}$, based on the~experimental data reported by~Soderstrom~\textit{et~al.}~\cite{Soderstrom_2011}. $f_\mathrm{d}$~in~\cref{eq:drop_heat_gma} is the~frequency of droplet detachment, and is defined as:

\begin{equation}
	f_\mathrm{d} = \frac{3 u_\mathrm{w} r_\mathrm{w}^2}{4 r_\mathrm{d}^3},
	\label{eq:drop_frequency_gma}
\end{equation}

\noindent
where, $u_\mathrm{w}$~is the~wire feed rate and $r_\mathrm{w}$~is~the~radius of the~welding wire. For metal transfer in the~spray mode, the~radius of the~molten metal droplets and the~welding wire are assumed to be the~same. Accordingly, the~magnitude of molten metal droplet velocity $u_\mathrm{d}$ just after detachment was approximated using the~correlation proposed by~Lin~\textit{et~al.}~\cite{Lin_2001}:

\begin{equation}
	u_\mathrm{d} = \frac{I}{2 \pi r_\mathrm{d}} \sqrt{{\frac{3 \mu_0}{\rho_\mathrm{d}}}} G,
	\label{eq:drop_velocity_gma}
\end{equation}

\noindent
where, $I$~is in Ampere, $r_\mathrm{d}$~the~radius of the~droplet in meters, $\mu_0$~the~vacuum permeability in~$\si{\henry\per\meter}$, $\rho_\mathrm{d}$~the~density of the~molten droplet in $\si{\kilogram\per\meter\cubed}$, and $G$~a~dimensionless constant introduced to obtain agreement with experimental measurements equal to 0.98 for steel electrodes.

The~process voltage~$U$ was assumed to be a~function of welding current and arc length~\cite{Lancaster_1986,Zhang_2018,Zhang_2020}, and was determined as follows:

\begin{equation}
	U = U_\mathrm{w} + U_\mathrm{o} + U_\mathrm{a}.
	\label{eq:voltage_variations_gma}
\end{equation}

\noindent
Here,~$U_\mathrm{w}$ is the~wire voltage assumed to be constant and equal to~$\SI{7}{\volt}$~\cite{Zhang_2018}, $U_\mathrm{o}$~the~sum of the~electrode fall voltages is assumed to be a~function of welding current $I$:

\begin{equation}
	U_\mathrm{o} = C_\mathrm{I} I + 10,
	\label{eq:fall_voltage_gma}
\end{equation}

\noindent
with $I$~in Ampere and $C_\mathrm{I}$~the~coefficient of variation of the~electric fall voltage with current equal to~$\SI{0.016}{\volt\per\ampere}$~\cite{Zhang_2018,Zhang_2020}. $U_\mathrm{a}$~in \cref{eq:voltage_variations_gma}~is the~arc column voltage:

\begin{equation}
	U_\mathrm{a} = C_\mathrm{e} \ell,
	\label{eq:column_voltage_gma}
\end{equation}

\noindent
with $\ell$~in meters and $C_\mathrm{e}$~the~electric field strength equal to~$\SI{1.09}{\volt\per\milli\meter}$~\cite{Zhang_2018,Zhang_2020}. Using the~data reported by~Tsai~and~Eagar~\cite{Tsai_1985}, the~distribution parameter~$\sigma_\mathrm{q}$ (in meters) was determined as follows:

\begin{equation}
	\sigma_\mathrm{q} = \SI{1.61e-1}{} \ell^{0.976} + \SI{2.23e-4}{} I^{0.395},
	\label{eq:dist_heat_gma}
\end{equation}

\noindent
with~$\ell$ in meters and~$I$ in Ampere. The~adjustment factor~$\mathscr{F}_\mathrm{q}$ was used to negate changes in the~total heat input due to surface deformations~\cite{Ebrahimi_2020,Ebrahimi_2019_conf}, which is defined as follows:

\begin{equation}
	\mathscr{F}_\mathrm{q} = \frac{\eta I U}{\iiint \limits_{\forall} S_\mathrm{q} \mathop{}\!\mathrm{d}V}.
	\label{eq:adj_heat_gma}
\end{equation}

\noindent
It should be noted that the~source term~$S_\mathrm{q}$ is only applied to the~top surface of the~workpiece.

The~sink term~$S_\mathrm{l}$ was added to the~energy equation to account for heat losses due to convection and radiation, and is determined as follows:

\begin{equation}
	S_\mathrm{l} = - \left[\mathscr{h}_\mathrm{c} \left(T - {T_\mathrm{0}}\right) + \mathscr{K}_\mathrm{b} \mathscr{E} \left(T^4 - {T_\mathrm{0}}^4\right)\right] \lVert \nabla\phi \rVert \frac{2\, \rho \, c_\mathrm{p}}{(\rho \, c_\mathrm{p})_\mathrm{m} + (\rho \, c_\mathrm{p})_\mathrm{g}},
	\label{eq:heat_loss_gma}
\end{equation}

\noindent
where, $\mathscr{h}_\mathrm{c}$~is the~heat transfer coefficient equal to~$\SI{25}{\watt\per\square\meter\per\kelvin}$~\cite{Johnson_2017}, $\mathscr{K}_\mathrm{b}$ the~Stefan--Boltzmann constant, and $\mathscr{E}$~the~radiation emissivity equal to~$0.45$~\cite{Sridharan_2011}.

\FloatBarrier

\subsection{Numerical implementation}
\label{sec:num_proc}	

The~computational model employed in the~present work was developed within the~framework of a~proprietary finite-volume solver, ANSYS Fluent~\cite{Ansys_192}. To implement the~source terms in the~governing equations and the~surface tension model, user-defined subroutines programmed in the~C programming language were used. The~computational domain contains about~$\SI{2.7e6}{}$ non-uniform hexahedral cells, with the~smallest cell spacing being set to~$\SI{80}{\micro\meter}$ in the~melt-pool region, which is sufficiently fine to obtain grid-independent solutions~\cite{Ebrahimi_2019_conf,Ebrahimi_2020,Ebrahimi_2021,Ebrahimi_2021_b}. The~cell spacing increases gradually from the~melt-pool region towards the~boundaries of the~computational domain and the~maximum cell size was limited to~$\SI{400}{\micro\meter}$. The~central-differencing scheme with second-order accuracy was employed for spatial discretisation of momentum advection and diffusive fluxes. A~first-order implicit scheme was employed for the~time marching, and a~fixed time-step size of~$\SI{2e-5}{\second}$ was used to keep the~value of the~Courant number $(\mathrm{Co} = \lVert \mathbf{u} \rVert \Delta t / \Delta x)$ below 0.25. To formulate the~advection of the~volume-fraction scalar field, an~explicit compressive VOF method~\cite{Ubbink_1997} was employed. Moreover, the~PRESTO~(pressure staggering option) scheme~\cite{Patankar_1980} and the~PISO~(pressure-implicit with splitting of operators) scheme~\cite{Issa_1986} was employed for the~pressure interpolation and coupling velocity and pressure fields, respectively. Simulations were executed in parallel on a~high-performance computing cluster, each on 70~cores (AMD~EPYC~7452) and the~total run-time was about~$\SI{290}{\hour}$.

\FloatBarrier

\subsection{Experimental setup and procedure}
\label{sec:exp_proc}
\begin{figure}[!htb] %[H] %[htbp]
	\centering
	\includegraphics[width=\linewidth]{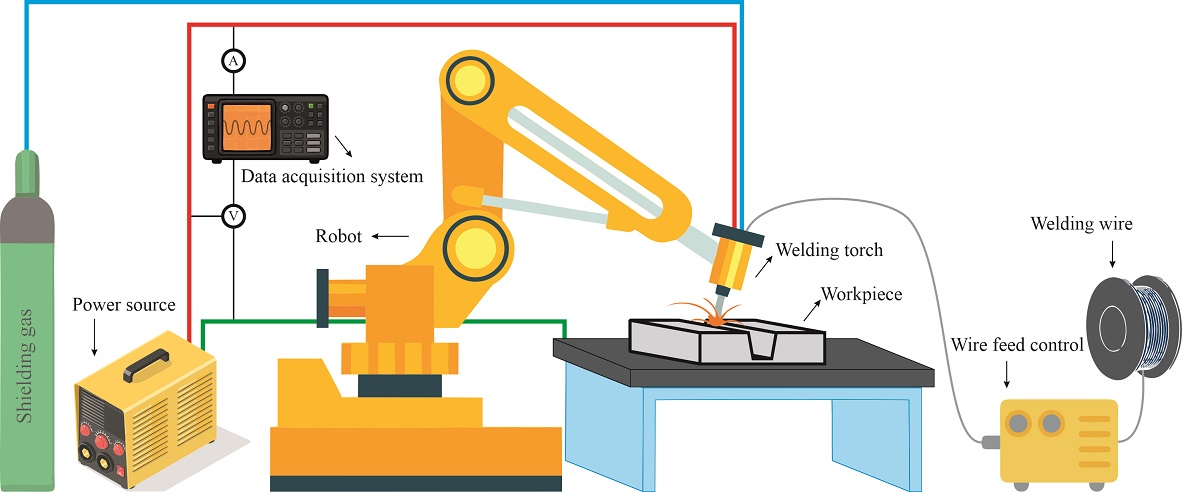}
	\caption{Schematic of the~experimental setup utilised in the~present work.}
	\label{fig:experimental_setup}
\end{figure}

The~general process parameters studied in the~present work are introduced in \cref{sec:problem_des}. \Cref{fig:experimental_setup} shows a~schematic drawing of the~experimental setup utilised in the~present work. A~Fronius~CMT~5000i power source that was attached to a~six-axis Fanuc robot was employed. Weld beads with a~length of $\SI{80}{\milli\meter}$ were deposited on the~workpiece with pre-machined grooves. Each experiment was repeated at least three times to ensure repeatability of the~tests. The~filler metal and the~workpiece employed in the~experiments were AISI~316L. Welding current and voltage were measured and recorded at a~frequency of $\SI{5}{\kilo\hertz}$ during the~experiments using a~Triton~4000 data acquisition system. Samples were cut after the~experiments to extract transverse cross-sections. The~cut samples were mounted and surface ground using silicon carbide (\ch{SiC}) papers with grit sizes varying from 80 to 2000 grit. Finally, the~samples were polished using colloidal alumina with particle sizes of $\SI{3}{\micro\meter}$ and $\SI{1}{\micro\meter}$ respectively. Fusion zones were revealed by chemical etching with Kallings Reagent I ($\SI{2}{\gram}$ \ch{CuCl2} + $\SI{40}{\milli\litre}$ \ch{HCl} + $\SI{40}{\milli\litre}$ \ch{C2H5OH} + $\SI{40}{\milli\litre}$ \ch{H2O}) for $\SI{3}{\second}$. Macrographs of the~fusion zones in the~etched specimens were obtained using a~Keyence digital microscope.

\FloatBarrier

\section{Results and discussion}
\label{sec:results}

\subsection{Model validation}
\label{sec:validation}
\begin{figure}[!htb] %[H] %[htbp]
	\centering
	\includegraphics[width=\linewidth]{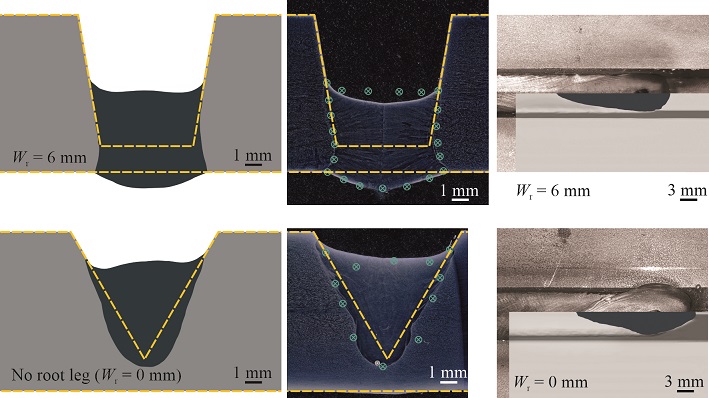}
	\caption{Comparison of the~melt-pool shapes obtained from the~present numerical simulations with experimental measurements for different groove shapes with a~welding current of $\SI{280}{\ampere}$ and a~travel speed of $\SI{7.5}{\milli\meter\per\second}$. Regions shaded in dark grey show the~melt-pool shape obtained from numerical simulations. The~computational cells that, at any stage during the~transient calculations of the~melting and re-solidification process, contained molten metal were marked to visualise the~melt-pool shape on the~transversal cross-section. Green symbols on experimental data show the~melt-pool boundary obtained from numerical simulations. Yellow dashed-lines indicate the~joint shape before welding.}
	\label{fig:validation}
\end{figure}

The~reliability and accuracy of the~present numerical predictions are benchmarked against experimentally measured melt-pool shapes. In this study, gas metal arc welding of workpieces with different groove shapes are considered, with a~welding current of $\SI{280}{\ampere}$ and a~travel speed of $\SI{7.5}{\milli\meter\per\second}$. \Cref{fig:validation} shows a~comparison between the~numerically predicted melt-pool shapes with those obtained from experiments for different groove shapes. The~computational cells containing molten metal were marked during the~calculation to visualise the~melt-pool shapes. It is worth noting that the~experiments were conducted after the~numerical simulations, which means no calibration is performed to tune the~numerical results. The~results indicate a~reasonable agreement between numerically predicted and experimentally measured melt-pool shapes. The~maximum deviation between the~predicted melt-pool dimensions and experimental measurements is found to be less than 10\%, demonstrating the~validity of the~present numerical simulations. This deviation might be caused by uncertainties associated with the~models employed to approximate the~temperature-dependent material properties at elevated temperatures, the~simplifying assumptions made to develop the~computational model such as those employed to determine droplet size, velocity and temperature, and uncertainties in determining the~boundary conditions in the~model.

\FloatBarrier
	
\subsection{Thermal and fluid flow fields}
\label{sec:thermal_fluid}
Once the~arc ignites and the~process begins, the~welding wire heats up to the~melting temperature and molten metal droplets form at the~wire tip that detach periodically from the~wire and deposit on the~workpiece surface as shown schematically in \cref{fig:schematic}. The~frequency of droplet detachment is directly proportional to the~wire feed rate and ranges between $\SI{147}{\hertz}$ and $\SI{187}{\hertz}$ for the~welding process parameters studied in the~present work (see~\cref{tab:welding_parameters}). To simplify the~numerical simulations and as described in \cref{sec:math_model}, the~filler metal droplets are assumed to be spherical and are incorporated into the~simulations with predefined velocity and temperature, which is a~common practice in modelling melt-pool behaviour in gas metal arc welding (see for instance, \cite{Zong_2021,Hu_2021,Hu_2008_b,Cho_2013}). The~qualitative melt-pool behaviour was found to be similar for different welding currents studied in the~present work. Therefore, representative results for the~cases with welding current of $\SI{220}{\ampere}$ are shown and discussed in the~paper. % and other data are presented in the~supplementary materials

\begin{figure}[!htb] %[H] %[htbp]
	\centering
	\includegraphics[width=\linewidth]{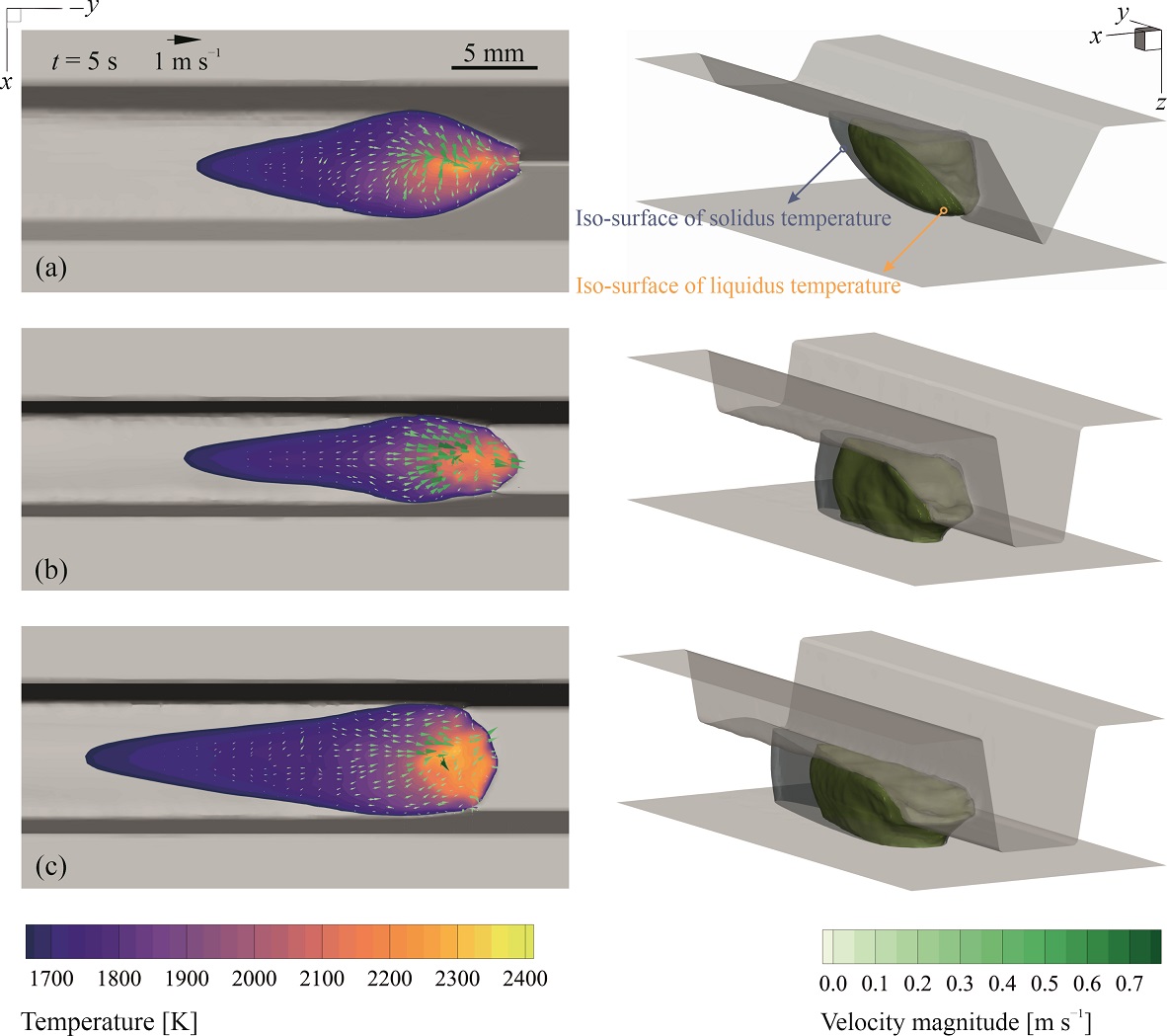}
	\caption{The~numerically predicted thermal and fluid fields over the~melt-pool surface (left column) and the~corresponding pool shape (right column) for different joint shapes at $t = \SI{5}{\second}$. (a) groove angle $\theta = 60^\circ$ and no root-leg ($W_\mathrm{r} = \SI{0}{\milli\meter}$), (b)~$\theta = 20^\circ$ and root-leg width $W_\mathrm{r} = \SI{4}{\milli\meter}$ and (c) $\theta = 20^\circ$ and root-leg width $W_\mathrm{r} = \SI{6}{\milli\meter}$. Wire feed rate $u_\mathrm{w} = \SI{7}{\meter\per\minute}$, welding current $I = \SI{220}{\ampere}$, and travel speed $V = \SI{7.5}{\meter\per\second}$. The~area between iso-surfaces of solidus and liquidus temperature shows the~mushy region.}
	\label{fig:temperature_velocity}
\end{figure}

The~thermal energy input from the~plasma arc in addition to the~thermal energy transported by the~molten metal droplets result in the~formation of a~melt pool. For the~process parameters studied in the~present work (see~\cref{tab:welding_parameters}), the~melt pool grows over time and reaches a~quasi-steady-state condition after about $\SI{3}{\second}$. \Cref{fig:temperature_velocity} shows a~partial view of the~workpiece encompassing the~melt pool and the~corresponding thermal and fluid flow fields over the~melt-pool surface for different groove shapes at $t = \SI{5}{\second}$ with wire feed rate $u_\mathrm{w} = \SI{7}{\meter\per\minute}$ and welding current $I = \SI{220}{\ampere}$. For the~cases shown in \cref{fig:temperature_velocity}, the~maximum surface temperature is less than $\SI{2310}{\kelvin}$ and the~value of the~temperature gradient of surface tension ($\partial \gamma / \partial T$) is mostly positive (see~\cref{fig:materials_properties}(e)). Hence, the~molten metal moves from the~cold area close to the~melt-pool rim towards the~hot central region, primarily due to the~Marangoni shear force induced over the~surface. Molten metal streams from the~melt-pool rim collide in the~central region and form a~complex unsteady asymmetric flow pattern in the~pool. A~similar flow pattern is observed experimentally in previous independent studies conducted by~\mbox{Wu~\textit{et~al.}~\cite{Wu_2021} and Zhao~\textit{et~al.}~\cite{Zhao_2009}}. The~maximum local molten metal velocity is about $0.7-\SI{0.8}{\meter\per\second}$ and corresponds to a~P\'eclet number ($\mathrm{Pe} = \rho c_\mathrm{p} \mathscr{D} \lVert \mathbf{u} \rVert / k$) larger than unity ($\mathscr{O}(400)$), which signifies that advection dominates the~energy transfer in the~melt pool and that the~process cannot be described adequately using a~thermal model without considering fluid flow. 

The~results suggest that the~energy transported to the~surrounding solid material markedly affects the~melt-pool shape. Although the~total heat input to the~material is the~same for the~cases shown in \cref{fig:temperature_velocity}, the~melt-pool shapes differ notably for different groove shapes. It appears that increasing the~width of the~root-leg results in a~decrease in the~amount of heat diffused to the~side walls of the~groove as the~height of the~deposit layer reduces, leading to an~increase in the~length of the~melt-pool as well as the~mushy-zone (\textit{i.e.} regions between the~solidus and liquid iso-surfaces in~\cref{fig:temperature_velocity}). Moreover, the~average fluid velocity in the~pool decreases with increasing width of the~root-leg, which can be attributed to the~decrease in the~magnitude of temperature gradients generated over the~surface. Among all the~cases studied in the~present work, full-penetration is observed only for those with root-leg, even for the~case with welding current $I = \SI{280}{\ampere}$. Evidently, a~higher welding current or a~lower travel speed is required to achieve full penetration using grooves without root-leg (\textit{i.e.} V-groove). However, increasing the~welding current or reducing the~travel speed results in an~increase in total heat input to the~material, which is often undesirable as it decreases the~cooling rate and can adversely affect the~properties of the~joint, particularly when austenitic stainless steels are used~\cite{Kumar_2011,Unnikrishnan_2014,Mohammed_2017}. Moreover, increasing the~welding current can lead to a~significant increase in arc force as the~arc force is proportional to the~welding current squared ($\lVert \mathbf{F}_\mathrm{arc} \rVert \propto I^2$)~\cite{Lancaster_1986}, and thus limiting the~welding current tolerance to avoid defects such as burn-through. Despite the~fact that employing a~root-leg can reduce the~welding current required to achieve full penetration, employing a~relatively wide root-leg may increase the~number of welding passes required to fill the~groove. 

\begin{figure}[!htb] %[H] %[htbp]
	\centering
	\includegraphics[width=\linewidth]{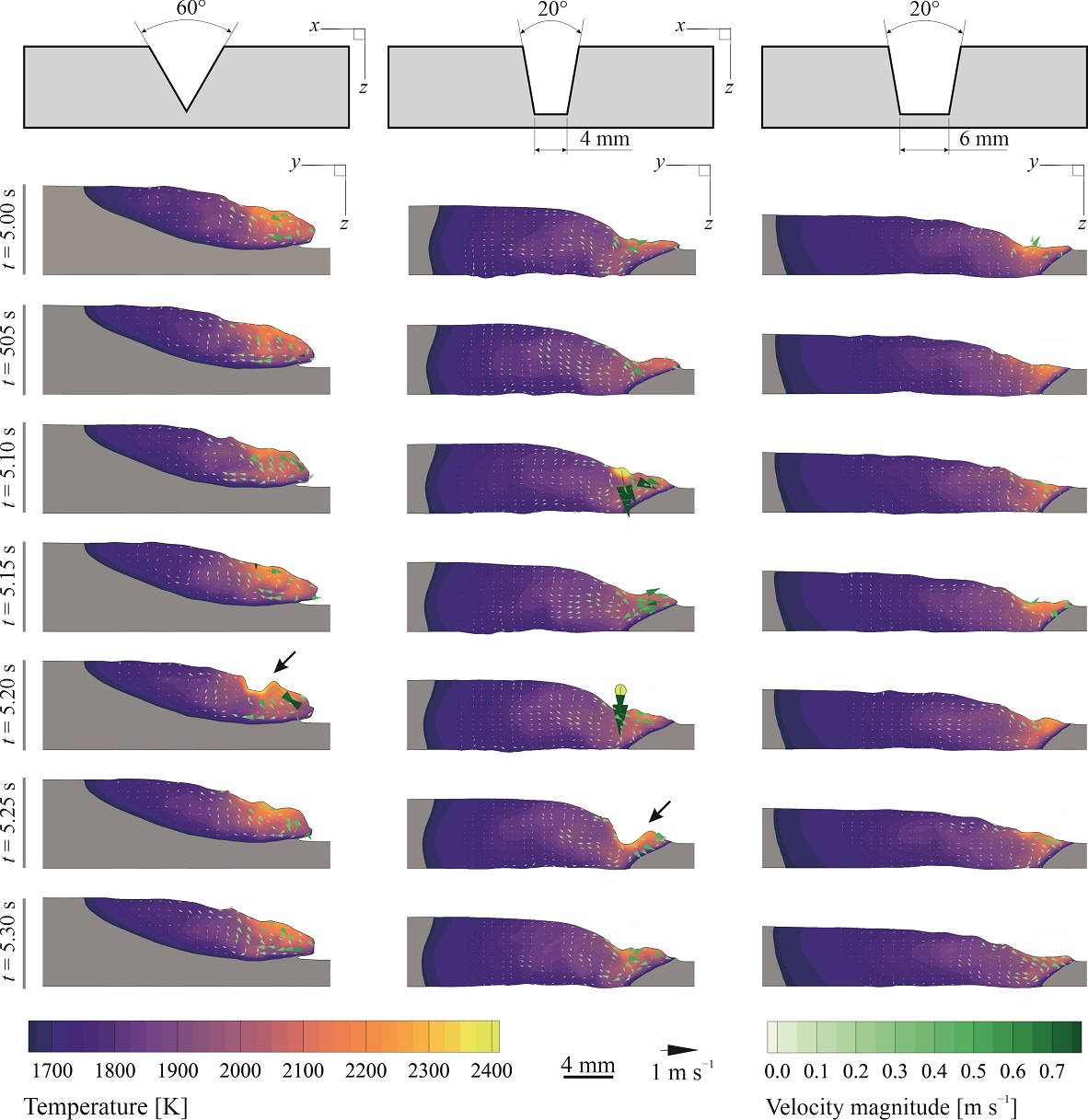}
	\caption{Melt-pool shape, temperature profile and velocity vectors in the~$x = 0$ plane for different joint shapes and time instances. (left column) groove angle $\theta = 60^\circ$ and no root-leg ($W_\mathrm{r} = \SI{0}{\milli\meter}$), (middle column)~$\theta = 20^\circ$ and root-leg width $W_\mathrm{r} = \SI{4}{\milli\meter}$ and (right column) $\theta = 20^\circ$ and root-leg width $W_\mathrm{r} = \SI{6}{\milli\meter}$. Wire feed rate $u_\mathrm{w} = \SI{7}{\meter\per\minute}$, welding current $I = \SI{220}{\ampere}$, and travel speed $V = \SI{7.5}{\meter\per\second}$.}
	\label{fig:meltpool_droplet_interaction}
\end{figure}

\Cref{fig:meltpool_droplet_interaction} shows thermal and fluid flow fields in the~$x = 0$ plane for different joint shapes and time instances. The~impingement of molten metal droplets on the~surface disturbs the~thermal and fluid flow field in the~pool and results in the~formation of a~crater and a~travelling wave over the~melt-pool surface, as indicated by arrows in \cref{fig:meltpool_droplet_interaction}. Moreover, the~periodic molten metal droplet impingement on the~melt pool enhances mixing in the~melt pool. The~molten metal droplet temperature ($T_\mathrm{d} = \SI{2500}{\kelvin}$) is above the~critical temperature at which the~sign of surface-tension temperature coefficient ($\partial \gamma / \partial T$) changes from positive to negative ($T_\mathrm{cr} \approx \SI{2250}{\kelvin}$); therefore, an~outward fluid flow is induced on the~surface in the~region where the~droplet is impinged due to Marangoni shear force. Soon after the~droplet is merged with the~melt pool, the~crater closes due to surface tension and hydrostatic forces, and the~surface temperature decreases to values less than $\SI{2310}{\kelvin}$ for which the~value of $\partial \gamma / \partial T$ is mostly positive. The~wave crests move radially outward towards the~met-pool rim and are reflected by the~solid edges of the~pool. Interactions between the~primary and reflected waves as well as the~forces acting on the~molten material result in complex melt-pool surface deformations and oscillations, as shown in~\cref{fig:temperature_velocity}. For the~cases studied in the~present work, the~frequency of the~droplet transfer in relatively high ($\mathscr{O}(\SI{170}{\hertz})$) and the~droplet sizes are relatively small compared to the~melt-pool dimension, resulting in a~smooth weld bead with negligible ripple formation.

\FloatBarrier

\section{Conclusion}
\label{sec:conclusions}
Three-dimensional numerical simulations were performed to systematically investigate the~effect of groove shape on melt-pool behaviour in root pass gas metal arc welding (GMAW). The~effects of melt-pool surface deformations on power-density distribution and the~forces applied to the~molten material were accounted for in the~present computational model. These effects are often neglected in numerical simulations of melt-pool behaviour in arc welding. Thermal and fluid flow fields in the~melt pool are visualised and described for different groove shapes. Moreover, experiments were conducted to validate the~numerical predictions.

Energy transfer in melt pools during gas metal arc welding is dominated by convection and thus thermal models without considering fluid flow cannot predict and describe the~melt-pool shape with sufficient accuracy. The~periodic impingement of molten metal droplets disturbs the~thermal and fluid flow fields in the~pool, resulting in an~even more complex flow pattern. For the~process parameters studied in the~present work, full-penetration was observed only for the~grooves with root-leg. Changes in the~groove shape have an~insignificant influence on the~flow pattern over the~surface, however the~groove shape affects the~energy transfer to the~surrounding solid material and thus alters the~melt-pool shape and can affect the~properties of the~joint. The~groove shape also affects the~melt-pool oscillatory behaviour as it influences the~reflection of the~waves generated due to the~molten metal droplet impingement. Moreover, the~groove shape can affect the~process window, which can be explored using the~simulation-based approach described in the~present work.

\section*{Acknowledgement}
\label{sec:acknowledgement}

This research was carried out under project number F31.7.13504 in the framework of the Partnership Program of the Materials innovation institute M2i (www.m2i.nl) and the Foundation for Fundamental Research on Matter (FOM) (www.fom.nl), which is part of the Netherlands Organisation for Scientific Research (www.nwo.nl). The~authors would like to thank the industrial partner in this project “Allseas Engineering B.V.” for the financial support.

\section*{Author Contributions}
\label{sec:author_contributions}

Conceptualisation, A.E. and I.M.R.; methodology, A.E.; software, A.E.; validation, A.E.; formal analysis, A.E.; investigation, A.E.; resources, A.E. and M.J.M.H.; data~curation, A.E. and A.B.; writing---original draft preparation, A.E.; writing---review and editing, A.E., A.B., C.R.K., M.J.M.H. and~I.M.R; visualisation, A.E.; supervision, C.R.K. and I.M.R.; project administration, A.E.; and funding acquisition, I.M.R. and M.J.M.H.

\section*{Conflict of interest}
\label{sec:conflict_of_interest}

The authors declare no conflict of interest.

\section*{Data availability}
\label{sec:data_availability}

The data generated in this study are available on reasonable request from the corresponding author.

\small{
	\bibliographystyle{elsarticle-num}
	\bibliography{ref}
}

\end{document}